\documentstyle[12pt]{ioplppt}                % use this for journal style

\begin{document}

\jl{1}

\title{Extension of the discrete KP hierarchy}

\author{A K Svinin\ftnote{1}{E-mail address : svinin@icc.ru}}

\address{Institute of System Dynamics and Control Theory,
Siberian Branch of Russian Academy of Sciences,
P.O. Box 1233, 664033 Irkutsk, Russia}

\begin{abstract}
We introduce the discrete hierarchy which naturally generalizes
well known discrete KP hierarchy.
\end{abstract}

\maketitle

\section{Introduction}

The main purpose of this paper is to show how the well known discrete
KP hierarchy \cite{ueno} can be generalized by adding additional multi-times.
What we get in results is referred to as extended discrete KP while
any subsystem of it corresponding to multi-time $t^{(n)} \equiv  (t^{(n)}_1 \equiv
x^{(n)}, t^{(n)}_2,...)$ we call $n$th discrete KP hierarchy.

We show in the paper that $n$th  discrete KP in fact is equivalent to
bi-infinite sequence of copies of differential KP hierarchy whose
Lax operators are connected with each other by compatible `gauge'
transformations. The compatibility of the latter turned out to be equivalent
to equations of motion which represent the first flow in $n$th discrete KP.

The paper is organized as follows. After giving some notations in Section 2,
in Section 3 we introduce and discuss the extension of the discrete KP.
Section 4 is devoted to provide relationship between $n$th discrete KP
and sequence of differential KP.

\section{The differential and discrete KP hierarchy}

Let us recall some basic facts about the differential KP hierarchy
in the spirit
of Sato theory \cite{date}, \cite{jimbo}, \cite{ohta}. This approach essentially
is based on the calculus of the pseudo-differential operators ($\Psi$DO's)
\cite{dickey2}.
For reasons of completeness a certain amount of notation has to be introduced.

The unknown functions (fields) depend on spatial variable
$t_1\equiv x\in{\bf R}^1$ and some evolution parameters $t_2, t_3,...$
The symbols $\partial$ and $\partial_p$ stand for derivation with respect
to $x$ and $t_p$, respectively. In what follows the symbol $t$ denotes
KP multi-time, i.e. infinite set of evolution parameters $(t_1, t_2, t_3,...)$.
Let $R$ be a commutative ring consisting of all smooth functions $a=a(x)$.
Then noncommutative ring $R[\partial, \partial^{-1})$ of $\Psi$DO's consists
of all formal expressions
\[
A = \sum_{i=-\infty}^{N}a_i(x)\partial^i,\;\;
N\in{\bf Z}
\]
with coefficients in $R$. One says that $\Psi$DO $A$ is of order $N$.
The operator $\partial : R\rightarrow R$ is entirely defined by generalized
Leibniz rule
\[
\partial^i\circ a =
\sum_{j=0}^{\infty}\left(i \atop j \right)a^{(j)}\partial^{i-j}
\]
where $a^{(j)}\equiv \partial^j a$. The adjoint of $A$ is given by
\[
A^{*} = \sum_{i=-\infty}^{N}(-\partial)^i\circ a_i.
\]
The important r\^ole in the theory plays decomposition of elements of
$R[\partial, \partial^{-1})$ into positive (differential) and negative (integral)
parts. We denote
\[
A_{+} = \sum_{i\geq 0}a_i(x)\partial^i,\;\;
A_{-} = \sum_{i\leq -1}a_i(x)\partial^i.
\]
respectively.

It is convenient to introduce a formal dressing operator $\hat{w} = 1
+ \sum_{k\in{\bf N}}w_k\partial^{-k}$. Then the KP
hierarchy can be represented via Sato--Wilson equations
\begin{equation}
\partial_p\hat{w} = - (\hat{w}\partial^p\hat{w}^{-1})_{-}\hat{w} =
(\hat{w}\partial^p\hat{w}^{-1})_{+}\hat{w} - \hat{w}\partial^p
\label{sato}
\end{equation}
or equivalently as Lax equations
\begin{equation}
\partial_p{\cal Q} = [({\cal Q}^p)_{+}, {\cal Q}] \equiv
({\cal Q}^p)_{+}{\cal Q} - {\cal Q}({\cal Q}^p)_{+}.
\label{lax0}
\end{equation}
on first-order $\Psi$DO ${\cal Q} = \hat{w}\partial\hat{w}^{-1} = \partial +
\sum_{k=1}^{\infty}u_k(t)\partial^{-k}$. The very important
observation is that evolution equations of the KP hierarchy are solved
in terms of single $\tau$-function satisfying an infinite set of bilinear
equations which are encoded in the fundamental bilinear identity
\begin{equation}
{\rm res}_z[\psi(t, z)\psi^{*}(t^{'}, z)] \equiv
\frac{1}{2\pi i}\oint_{0}\psi(t, z)\psi^{*}(t^{'}, z)dz = 0.
\label{bi}
\end{equation}
Recall that formal Backer -- Akhiezer wave function $\psi$ and its
conjugate  $\psi^{*}$ entering fundamental identity are related to KP
$\tau$-function via
\[
\psi(t, z) =
\frac{\tau(t - [z^{-1}])}
{\tau(t)}\exp(\xi(t, z)),\;\;
\psi^{*}(t, z) =
\frac{\tau(t + [z^{-1}])}
{\tau(t)}\exp(-\xi(t, z))
\]
with $\xi(t, z) = \sum_{p=1}^{\infty}t_pz^p$ and
$[z^{-1}] = (1/z, 1/(2z^2), 1/(3z^3),...)$. Then the bilinear
identity (\ref{bi}) becomes
\[
\sum_{k=0}^{\infty}p_k(-2a)p_{k+1}(\tilde{D}_t)\tau\circ\tau = 0,\;\;\;
\forall a=(a_1, a_2,...).
\]
A few remarks are in order.
For given polynomial
$p(\partial/\partial t_1, \partial/\partial t_2,...)$ in
$\partial/\partial t_i$, one defines
\[
p(D_{t_1}, D_{t_2},...)f\circ g
\]
\[
\left.
= p\left(\frac{\partial}{\partial u_1}, \frac{\partial}{\partial u_2},...\right)
f(t_1+u_1, t_2+u_2,...)g(t_1-u_1, t_2-u_2,...)\right|_{u=0}.
\]
In what follows $\tilde{D}_t\equiv (D_{t_1}, {1 \over 2}D_{t_2},
{1 \over 3}D_{t_3},...).$
It is worth also to recall following identity\footnote{here $t = (t_1, t_2,...)$}.:
\begin{equation}
\left.
\frac{1}{k!}\left(\frac{d}{du}\right)^kf(t+[u])g(t-[u])\right|_{u=0} =
p_k(\tilde{D})f\circ g
\label{identity}
\end{equation}
which will be useful in the following. Here Schur polynomials $p_k(t)$ are
defined through
\[
\exp\left(\sum_{p=1}^{\infty}t_pz^p\right) = \sum_{k=0}^{\infty}
z^kp_k(t).
\]

The discrete KP is a KP hierarchy where continuous space variable gets
replaced by a discrete $i$-variable. More exactly, equations of motion
of discrete KP are encoded by Lax equation
\[
\frac{\partial Q}{\partial t_p} = [Q_{+}^p, Q]
\]
on difference operator
$
Q = \Lambda + \sum_{k\in{\bf N}}a_{k-1}\Lambda^{1-k}.
$
The main reference in this context is the
paper of Ueno and Takasaki \cite{ueno}. This hierarchy as well
as a large class of its solutions is well described in \cite{adler1}.
What we learn from this work (see also \cite{bonora}) is that the
discrete KP is tantamount to bi-infinite sequence of differential KP copies
`glued' together by Darboux-B\"acklund (DB) transformations. This
leads to certain bilinear relations connecting the consecutive
KP $\tau$-functions. It should be noted that integrable systems
which are chains of infinitely many copies of KP-type differential
hierarchies turn out to be useful in matrix models \cite{aratyn},
\cite{dickey1}, \cite{bonora}. In the work \cite{dickey} by Dickey, it was shown
that the discrete KP hierarchy is the most natural generalization
of the modified KP.

\section{Extended discrete KP hierarchy}

Throughout the paper, we will be dealing with $\infty\times\infty$
matrices.
Given the shift operator $\Lambda = (\delta_{i,j-1})_{i,j\in{\bf Z}}$
and `spectral' parameter $z$ one considers the following spaces
of the difference operators\footnote{$z$ acts as component-wise multiplication.}:
\[
{\cal D}^{(n,r)} = \left\{
\sum_{-\infty\ll k<\infty}l_k z^{k(n-1)}\Lambda^{r-kn}\right\} = {\cal D}_{-}^{(n,r)} + {\cal D}_{+}^{(n,r)}
\]
with $l_k \equiv (l_k(i))_{i\in{\bf Z}}$ being bi-infinite diagonal
matrices. One can easily check, the following properties:
\[
{\cal D}^{(n,r_1)}\cdot{\cal D}^{(n,r_2)}\subset{\cal D}^{(n,r_1+r_2)},\;\;
\Lambda\cdot{\cal D}^{(n,r)}\subset{\cal D}^{(n,r+1)},\;\;
\]
\[
{\cal D}^{(n,r)}\cdot\Lambda\subset{\cal D}^{(n,r+1)},\;\;
z^{n-1}{\cal D}^{(n,r)}\subset{\cal D}^{(n,r+n)}.
\]

{\bf Remark 1.} In the case $n=1$, a dependence of ${\cal D}^{(1,r)}$ on $r$ make no
sense because $L\in{\cal D}^{(1,r)}$ does not depend on $z$.

The splitting of ${\cal D}^{(n,r)}$
into `negative' and `positive' parts is defined as follows:
\[
{\cal D}_{-}^{(n,r)} = \left\{
\sum_{r-kn\le -1}l_kz^{k(n-1)}\Lambda^{r-kn}\right\},\;\;
{\cal D}_{+}^{(n,r)} = \left\{
\sum_{r-kn\ge 0}l_kz^{k(n-1)}\Lambda^{r-kn}\right\}.
\]
In the following we assume that  the entries of $l_k$'s
may depend on multi-time $t\equiv(t_p^{(n)})_{p,n\in{\bf N}}$.
For corresponding time derivatives we use following notation:
$\partial_p^{(n)} = \partial/\partial t_p^{(n)}$
and $\partial^{(n)} = \partial/\partial x^{(n)}$, where
$x^{(n)}\equiv t_1^{(n)}$.

The phase space ${\cal M}$ consists of the entries of diagonal matrices
$w_k = (w_k(i))_{i\in{\bf Z}},\; k\in{\bf N}$. For each $n\in{\bf N}$,
we define the `wave' operator
\begin{equation}
S^{(n)} = I + \sum_{k\in{\bf N}}w_k z^{k(n-1)}\Lambda^{-kn}\in I +
{\cal D}_{-}^{(n,0)}
\label{S^n}
\end{equation}
and the corresponding Lax operator
\begin{equation}
Q^{(n)} \equiv S^{(n)}\Lambda S^{(n)-1} = \Lambda +
\sum_{k\in{\bf N}}a_{k-1}^{(n)} z^{k(n-1)}\Lambda^{1-kn}\in{\cal D}^{(n,1)}.
\label{LAX-op}
\end{equation}
It is clear, the coordinates $a_{k}^{(n)}$ are related with original ones
by some polynomial relations. For example, from (\ref{LAX-op}) one can read
off the following:
\[
a_0^{(n)}(i) = w_1(i) - w_1(i+1),
\]
\[
a_1^{(n)}(i) = w_2(i) - w_2(i+1) + w_1(i-n+1)(w_1(i+1) - w_1(i)),
\]
\[
a_2^{(n)}(i) = w_3(i) - w_3(i+1) + w_1(i-2n+1)(w_2(i+1) - w_2(i))
\]
\[
+ w_2(i-n+1)(w_1(i+1) - w_1(i))
\]
\[
+ w_1(i-2n+1)w_1(i-n+1)(w_1(i) - w_1(i+1)).
\]

Now we are in position to define the flows on ${\cal M}$ with respect
to parameters $t_p^{(n)}$. It will be managed by the equations
of motion on the `wave' operator
\footnote{henceforth we shall
employ the short-hand notation $S$ for $S^{(n)}$ and $Q$ for $Q^{(n)}$
whenever this will not lead to a confusion.}
\begin{equation}
\begin{array}{c}
\displaystyle
z^{p(n-1)}\frac{\partial S}{\partial t_p^{(n)}}  = Q_{+}^{pn}S - S\Lambda^{pn},\;\;
\in{\cal D}^{(n,pn)} \\[0.4cm]
\displaystyle
z^{p(n-1)}\frac{\partial (S^{-1})^T}{\partial t_p^{(n)}} = (S^{-1})^T\Lambda^{-pn} -
(Q_{+}^{pn})^T(S^{-1})^T.
\end{array}
\label{SW}
\end{equation}
We should perhaps note that the first and second equations in (\ref{SW})
in fact are equivalent. Evolutions of $S$ induces evolutions of
$Q$  in the form of the Lax equations
\begin{equation}
z^{p(n-1)}\frac{\partial Q}{\partial t_p^{(n)}} = [Q_{+}^{pn}, Q]\;\;
\in{\cal D}^{n,pn+1}.
\label{LAX}
\end{equation}
One can easily check that $[Q_{+}^{pn}, Q] =
- [Q_{-}^{pn}, Q]$ is of the same form as l.h.s.
of (\ref{LAX}) and therefore (\ref{LAX}) and equivalent equations
(\ref{SW}) are properly defined.

Obviously, the discrete KP hierarchy can be regarded
as a subsystem of (\ref{SW}) with respect to infinite set of parameters
$t^{(1)} = (t^{(1)}_1, t^{(1)}_2,...)$.
For this reason, we will refer to (\ref{SW}) as extended discrete
KP hierarchy. The subsystem of (\ref{SW}) corresponding to evolution
parameters $t^{(n)} = (t^{(n)}_1, t^{(n)}_2,...)$ we will call,
according to \cite{svinin}, $n$th discrete KP hierarchy.

In the following it will be useful also to consider evolution equations
\begin{equation}
z^{p(n-1)}\frac{\partial Q^r}{\partial t_p^{(n)}} = [Q_{+}^{pn}, Q^r],\;\;
r\in{\bf Z}
\label{LAX1}
\end{equation}
being consequences of (\ref{LAX}).
It is easy to see that $r$th power of $Q$ is of the form
\[
Q^r = \Lambda^r + \sum_{k\in{\bf N}}a_{k-1}^{(n,r)} z^{k(n-1)}\Lambda^{r-kn}\in{\cal D}^{(n,r)}.
\]
with diagonal matrices $a_k^{(n,r)}$ whose entries are polynomially expressed
via original coordinates $w_k(i)$.

\section{$n$th discrete KP}

Define $\chi(z) = (z^i)_{i\in{\bf Z}}$, $\chi^{*}(z) = \chi(z^{-1})$
and wave vectors
\begin{equation}
\Psi(t, z) = W\chi(z),\;\;
\Psi^{*}(t, z) = (W^{-1})^T\chi^{*}(z)
\label{wv}
\end{equation}
where $W\equiv S\exp(\sum_{p=1}^{\infty}t_p^{(n)}\Lambda^p)$.
Discrete linear system
\begin{equation}
\begin{array}{c}
\displaystyle
Q\Psi(t, z) = z\Psi(t, z),\;\;Q^T\Psi^{*}(t, z) = z\Psi^{*}(t, z), \\[0.4cm]
\displaystyle
z^{p(n-1)}\partial_p^{(n)}\Psi = Q_{+}^{pn}\Psi,\;\;
z^{p(n-1)}\partial_p^{(n)}\Psi^{*} = - (Q_{+}^{pn})^T\Psi^{*}
\end{array}
\label{DLS1}
\end{equation}
are evident consequence of (\ref{SW}) and (\ref{wv}).
>From (\ref{wv}), it follows
\[
\Psi_i(t, z) =
z^i(1 + w_1(i)z^{-1} + w_2(i)z^{-2} + ...)e^{\xi(t^{(n)}, z)}
\]
\[
= z^i(1 + w_1(i)\partial^{(n)-1} + w_2(i)\partial^{(n)-2} + ...)
e^{\xi(t^{(n)}, z)}
\]
\[
\equiv
z^i\hat{w}_i(\partial^{(n)})e^{\xi(t^{(n)}, z)} \equiv
z^i\psi_i(t, z).
\]

Next, we are going to show equivalence of
$n$th discrete KP to bi-infinite sequence of differential KP copies ``glued" together
by compatible gauge transformations one of which can be recognized
as DB transformation mapping ${\cal Q}_i \equiv \hat{w}_i\partial^{(n)}
\hat{w}_i^{-1}$ to ${\cal Q}_{i+n} \equiv \hat{w}_{i+n}\partial^{(n)}\hat{w}_{i+n}^{-1}$.

{\bf Proposition 1.} {\it The following three statements are equivalent:

(i) The wave vector $\Psi(t, z)$ satisfies discrete linear system
\begin{equation}
Q^r\Psi(t, z) = z^r\Psi(t, z),\;\;
z^{n-1}\partial^{(n)}\Psi = Q_{+}^n\Psi,\;\;
r\in{\bf Z};
\label{DLS2}
\end{equation}

(ii) The components $\psi_i$ of the vector $\psi \equiv (\psi_i =
z^{-i}\Psi_i)_{i\in{\bf Z}}$ satisfy
\begin{equation}
G_i^{(r)}\psi_i(t, z) = z\psi_{i+n-r}(t, z),\;\;
H_i\psi_i(t, z) = z\psi_{i+n}(t, z)
\label{system1}
\end{equation}
with $H_i\equiv \partial^{(n)} - \sum_{s=1}^na_0^{(n)}(i+s-1)$ and
\[
G_i^{(r)}\equiv H_i + a_0^{(n,r)}(i+n-r)
\]
\[
+ a_1^{(n,r)}(i+n-r)H_{i-n}^{-1} +
a_2^{(n,r)}(i+n-r)H_{i-2n}^{-1}H_{i-n}^{-1} + ... ;
\]

(iii) For sequence of $\partial^{(n)}$-dressing operators $\{\hat{w}_i,\;
i\in{\bf Z}\}$ the equations
\begin{equation}
G_i^{(r)}\hat{w}_i = \hat{w}_{i+n-r}\partial^{(n)},\;\;
H_i\hat{w}_i = \hat{w}_{i+n}\partial^{(n)}
\label{system2}
\end{equation}
hold.
}

{\bf Remark 2.} Since $Q^0 = I$ and $a_k^{(n,0)}(i)=0$, we have in this case
$G_i^{(0)} = H_i$.

{\bf Proof of Proposition 1.} Rewrite equations (\ref{DLS2}) in
explicit form
\[
\Psi_{i+r} + a_0^{(n,r)}(i)z^{n-1}\Psi_{i+r-n} +
a_1^{(n,r)}(i)z^{2(n-1)}\Psi_{i+r-2n} + ... = z^r\Psi_i,
\]
\[
z^{n-1}\partial^{(n)}\Psi_i = \Psi_{i+n} +
\left(\sum_{s=1}^na_0^{(n)}(i+s-1)\right)\Psi_i.
\]
In terms of wave functions $\psi_i$ the latter is rewritten as
\begin{equation}
z\psi_{i+r} + a_0^{(n,r)}(i)\psi_{i+r-n} +
\frac{1}{z}a_1^{(n,r)}(i)\psi_{i+r-2n} + ... = z\psi_i,
\label{first}
\end{equation}
\begin{equation}
\partial^{(n)}\psi_i = z\psi_{i+n} +
\left(\sum_{s=1}^na_0^{(n)}(i+s-1)\right)\psi_i.
\label{second}
\end{equation}
One sees that equation (\ref{second}) coincides with second one in
(\ref{system2}). Shifting $i\rightarrow i-r+n$ in (\ref{first})
and combining it with (\ref{second}) one can obtain first equation
in (\ref{system2}). Therefore we proved ${\rm (i)}\Rightarrow {\rm (ii)}$.
The converse also easily can be showed by returning to the functions $\Psi_i$.
The equivalence ${\rm (ii)}\Leftrightarrow {\rm (iii)}$ follows from representation
$\psi_i(t, z) = \hat{w}_ie^{\xi(t^{(n)}, z)}$. $\Box$

Let us write down in explicit form
equations of motion coded in Lax equation
\begin{equation}
z^{n-1}\partial^{(n)}Q^r = [Q_{+}^n, Q^r]
\label{partial}
\end{equation}
being consistency condition of linear discrete
system (\ref{DLS2}). We have
\begin{equation}
\eqalign{
\partial^{(n)}a_k^{(n,r)}(i) = a_{k+1}^{(n,r)}(i+n) - a_{k+1}^{(n,r)}(i) \\
\displaystyle
+ a_{k}^{(n,r)}(i)\left(
\sum_{s=1}^n a_0^{(n)}(i+s-1) - \sum_{s=1}^n a_0^{(n)}(i+s+r-(k+1)n-1)
\right), \\
k = 0, 1,...
}
\label{lattice}
\end{equation}
where $a_0^{(n,r)}(i) = \sum_{s=1}^r a_0^{(n)}(i+s-1)$ in the case
$r\in{\bf N}$ and $a_0^{(n,r)}(i) = - \sum_{s=1}^{-r} a_0^{(n)}(i-s)$
when $r\in -{\bf N}$. Notice that simplest form of these flows is
in original coordinates:
\[
\partial^{(n)}w_k(i) = w_{k+1}(i+n) - w_{k+1}(i) + w_k(i)(w_1(i) -
w_1(i+n)).
\]

The system (\ref{lattice})
allows for obvious reductions specified by conditions
$a_k^{(n,r)}(i)\equiv 0$ when $k\ge k_0$ with $k_0\in{\bf N}$.
Let us spend a few lines to list a collection of integrable differential-difference
systems known from literature which can be derived as reductions
of the general system (\ref{lattice}).

{\bf Remark 3.} In what follows, when we say that
given system coincides with that in some reference it means that
these systems are the same up to very simple --- `cosmetic' ---
transformations.

{\bf Remark 4.} Most part of examples below can be found in \cite{svinin1}.
Unfortunately the case $r\le -1$ in this work was overlooked (see example 4).

{\bf Example 1.} Consider the case $n=1, r=2, k_0=2$. Corresponding
reduction of (\ref{lattice}) reads\footnote{here and in what follows
$\phantom{q}^{\prime}\equiv \partial/\partial x^{(n)}$ with corresponding
$n$.}
\[
a_0^{\prime}(i) + a_0^{\prime}(i+1)
\]
\[
= (a_0(i) + a_0(i+1))(a_0(i) - a_0(i+1)) + a_1(i+1) - a_1(i),
\]
\[
a_1^{\prime}(i) = 0.
\]
Actually, we have in this case one-field lattice\footnote{For one-field
lattices we use notation $a_0(i)=r_i$}
\begin{equation}
r_i^{\prime} + r_{i+1}^{\prime} = r_i^2 - r_{i+1}^2 + \nu_i,
\label{10}
\end{equation}
with $\nu_i = a_1(i+1) - a_1(i)$ being some constants. As is known the
lattice (\ref{10}) describes elementary Darboux transformation for
Schr\"odinger operator $L = \partial^2 + q(x)$.
An interesting property of the lattice (\ref{10}) is that it
reduces to Painlev\'e transcedents $P_4$ and $P_5$ due to imposing
periodicity conditions
$
r_{i+N} = r_i,\;\; \nu_{i+N} = \nu_i
$
for $N=3$ and $N=4$, respectively \cite{adler}.

{\bf Example 2.}  In the case $n=1,\: r=1, k_0\ge 2$ we obtain well known
generalized Toda systems known also as Kupershmidt ones \cite{kuper}
\begin{equation}
\begin{array}{l}
\displaystyle
a_{0}^{\prime}(i) = a_1(i+1) - a_1(i), \\[0.4cm]
\displaystyle
a_{k}^{\prime}(i) = a_{k}(i)\left(a_0(i) - a_0(i-k)\right)  \\[0.4cm]
+ a_{k+1}(i+1)-a_{k+1}(i),\;\;
k = 1,..., k_0-1.
\end{array}
\label{toda}
\end{equation}
In particular if $k_0=2$ we obtain ordinary Toda lattice in
polynomial form
\begin{equation}
\begin{array}{l}
\displaystyle
a_{0}^{\prime}(i) = a_1(i+1) - a_1(i), \\[0.4cm]
\displaystyle
a_{1}^{\prime}(i) = a_{1}(i)\left(a_0(i) - a_0(i-1)\right).
\end{array}
\label{TODA}
\end{equation}
Defining $u_i$ by relation $a_0(i) = -u_i^{\prime}$ and
$a_1(i) = \exp(u_{i-1}-u_{i})$ we arrive at more familiar exponential
form of the Toda lattice $u_i^{\prime\prime} = e^{u_{i-1}-u_i} - e^{u_i-u_{i+1}}.$

{\bf Example 3.} Let $n\ge 2,\: r = 1,\: k_0 = 1$. This choice corresponds
to Bogoyavlenskii lattices \cite{bogoyavlenskii}
\begin{equation}
r_{i}^{\prime} = r_{i}\left(\sum_{s=1}^{n-1}r_{i+s} -
\sum_{s=1}^{n-1}r_{i-s}\right).
\label{volterra}
\end{equation}

{\bf Example 4.} In the case $n=1,\: r=-1,\: k_0=2$ we have Belov--Chaltikian
lattice \cite{belov}
\[
a_{0}^{\prime}(i) = a_1(i+1) - a_1(i+2) + a_{0}(i)\left(a_0(i+1) - a_0(i-1)\right),
\]
\[
a_{1}^{\prime}(i) = a_{1}(i)\left(a_0(i) - a_0(i-3)\right).
\]

{\bf Example 5.} In the case $n=1,\: r=2,\: k_0=3$ we have the system
\begin{equation}
\begin{array}{l}
a_0^{\prime}(i) + a_0^{\prime}(i+1) = a_0^2(i) - a_0^2(i+1) + a_1(i+1) - a_1(i), \\[0.4cm]
a_1^{\prime}(i) = a_2(i+1) - a_2(i), \\[0.4cm]
a_2^{\prime}(i) = a_2(i)(a_0(i) - a_0(i-1)).
\end{array}
\label{MB}
\end{equation}
As can be checked the Miura-like transformation
\[
a_0(i) = b(i+1) - b(i),\: a_1(i) = a(i),\: a_2(i) = \frac{c(i)}{c(i-1)}
\]
define a mapping of solutions of the system
\[
\begin{array}{l}
\displaystyle
a^{\prime}(i) = \frac{c(i+1)}{c(i)} - \frac{c(i)}{c(i-1)}, \\[0.4cm]
b^{\prime}(i) + b^{\prime}(i+1) = a(i) - (b(i+1) - b(i))^2, \\[0.4cm]
c^{\prime}(i) = c(i)(b(i+1) - b(i))
\end{array}
\]
which appears in \cite{wu} into solutions of the lattice (\ref{MB}).
As was observed in \cite{svinin1}, higher counterpart of (\ref{MB})
is the Blaszak--Marciniak lattice \cite{blaszak}.

Define an infinite set of
$\Psi$DO's $\{G_i^{(\ell, r)},\: i, \ell, r\in{\bf Z}\}$ by means
of the following recurrence relations:
\begin{equation}
\begin{array}{c}
G_i^{(\ell+1,r)} = G_{i+n}^{(\ell, r)}H_i,\;\;
\ell = 0, 1, 2,... \\[0.4cm]
G_i^{(\ell-1,r)} = G_{i-n}^{(\ell, r)}H_{i-n}^{-1},\;\;
\ell = 0, -1, -2,...
\end{array}
\label{recurrence}
\end{equation}
with
\[
G_i^{(0,r)}\equiv G_{i-n}^{(r)}H_{i-n}^{-1} =
1 + a_0^{(n,r)}(i-r)H_{i-n}^{-1}
\]
\[
+ a_1^{(n,r)}(i-r)H_{i-2n}^{-1}H_{i-n}^{-1} +
a_2^{(n,r)}(i-r)H_{i-3n}^{-1}H_{i-2n}^{-1}H_{i-n}^{-1} + ...
\]
It is important to observe that
\begin{equation}
a_{\ell}^{(n,r)}(i+\ell n-r) = {\rm res}_{\partial^{(n)}}G_i^{(\ell,r)}.
\label{res}
\end{equation}

{\bf Proposition 2.} {\it Following auxiliary equations hold:
\begin{equation}
G_{i}^{(\ell, r)}\psi_i = z^{\ell}\psi_{i+\ell n-r}.
\label{res-aux}
\end{equation}
}

{\bf Proof.} By induction. Let $\ell=0$, then
\[
G_i^{(0,r)}\psi_i = G_{i-n}^{(r)}H_{i-n}^{-1}\psi_i =
z^{-1}G_{i-n}^{(r)}\psi_{i-n} = \psi_{i-r}.
\]
Now suppose that (\ref{res-aux}) is true for some $l$, then
\[
G_i^{(\ell+1, r)}\psi_i = G_{i+n}^{(\ell, r)}H_i\psi_i =
zG_{i+n}^{(\ell, r)}\psi_{i+n} =
z^{\ell+1}\psi_{i+(\ell+1)n-r}.
\]
This proves (\ref{res-aux}) for positive integers $\ell$. The similar
arguments are used for negative $\ell$'s.
$\Box$

As consequence of the proposition, we obtain $G_i^{(\ell, \ell n)} =
{\cal Q}_i^{\ell}$.
Notice that the equation (\ref{res-aux}) in equivalent form is rewritten as
\begin{equation}
G_i^{(\ell, r)}\hat{w}_i = \hat{w}_{i+\ell n-r}\partial^{(n)\ell}.
\label{equiv}
\end{equation}

{\bf Proposition 3.} {\it The relation
\begin{equation}
G_{i+\ell_2n - r_2}^{(\ell_1, r_1)}G_i^{(\ell_2, r_2)} =
G_i^{(\ell_1 + \ell_2, r_1 + r_2)}
\label{holds}
\end{equation}
holds.}

{\bf Proof.} Taking into account (\ref{equiv}), we obtain that left
multiplication
of l.h.s. and r.h.s. of  (\ref{holds}) on $\hat{w}_i$ gives the same,
namely $\hat{w}_{i+(\ell_1+\ell_2)n-r_1-r_2}(\partial^{(n)})^{\ell_1+\ell_2}$.
This proves proposition. $\Box$

As we have mentioned above, consistency condition of linear system
(\ref{DLS2}) being expressed in the form of Lax equation reads in explicit form
as lattice (\ref{lattice}). As consequence of proposition 3 we
obtain that this system guarantee the validity of permutation relations
\begin{equation}
G_{i+\ell_2n-r_2}^{(\ell_1,r_1)} G_i^{(\ell_2,r_2)} =
G_{i+\ell_4n-r_4}^{(\ell_3,r_3)}G_i^{(\ell_4,r_4)}
\label{permutation}
\end{equation}
with arbitrary integers $\{\ell_k, r_k\}_{k=1}^4$ such that
$\ell_1 + \ell_2 = \ell_3 + \ell_4$ and $r_1 + r_2 = r_3 + r_4$.
It is clear that permutation relation (\ref{permutation}) can be extended
on that with arbitrary number of cofactors. In addition since $G_i^{(0,0)} =
1$ we have
\[
G_i^{(\ell, r)^{-1}} = G_{i+\ell n - r}^{(-\ell, -r)}
\]
>From the above we learn that the system (\ref{lattice}) guarantee
that the set of bi-infinite
sequences of $\Psi$DO's $\{G_i^{(\ell, r)},\: i\in{\bf Z}\}$
endowed with the multiplication rule (\ref{holds}) bears the structure of
the group isomorphic to ${\bf Z}\times{\bf Z}$.

{\bf Proposition 4.} {\it By virtue of (\ref{equiv}) and its consistency
condition (\ref{permutation}), $\partial^{(n)}$-Lax operators ${\cal Q}_i$
are connected with each other by invertible compatible gauge
transformations
\begin{equation}
{\cal Q}_{i+\ell n - r} = G_i^{(\ell, r)}{\cal Q}_iG_i^{(\ell, r)^{-1}}.
\label{similarity1}
\end{equation}
}

{\bf Remark 5.} Since ${\cal Q}_i^{\ell} = G_i^{(\ell, \ell n)}$, the relation
(\ref{similarity1}) in the case $r = \ell n$ becomes trivial identity.

{\bf Proof of proposition 4.} Taking into account (\ref{equiv}), we have
\[
{\cal Q}_{i+\ell n-r} = \hat{w}_{i+\ell n-r}\partial\hat{w}_{i+\ell n-r}^{-1} =
(G_i^{(\ell,r)}\hat{w}_{i}\partial^{(n)-1})\partial^{(n)}
(\partial^{(n)}\hat{w}_{i}^{-1}G_i^{(\ell,r)-1})
\]
\[
= G_i^{(\ell, r)}\hat{w}_{i}\partial^{(n)}\hat{w}_{i}^{-1}G_i^{(\ell,r)-1} =
G_i^{(\ell,r)}{\cal Q}_iG_i^{(\ell,r)-1}.
\]
The mapping ${\cal Q}_i\rightarrow\tilde{{\cal Q}}_i = {\cal Q}_{i+m}$,
where $m=\ell n-r$ we denote as $s_m$.

Let $m_1 = \ell_1n - r_1$  and $m_2 = \ell_2n - r_2$.
By virtue of (\ref{permutation}), where $\ell_3=\ell_2$, $\ell_4=\ell_1$,
$r_3=r_2$ and $r_4=r_1$ we get
\[
{\cal Q}_{i+m_1+m_2} =
G_{i+m_2}^{(\ell_1, r_1)}{\cal Q}_{i+m_2}G_{i+m_2}^{(\ell_1,r_1)-1}
\]
\[
= G_{i+m_2}^{(\ell_1,r_1)}G_{i}^{(r_2)}{\cal Q}_iG_{i}^{(\ell_2,r_2)-1}G_{i+m_2}^{(\ell_1,r_1)-1} =
G_{i+m_1}^{(\ell_2,r_2)}G_{i}^{(\ell_1,r_1)}{\cal Q}_iG_{i}^{(\ell,r_1)-1}G_{i+m_1}^{(\ell_2,r_2)-1}
\]
\[
= G_{i+m_1}^{(\ell_2,r_2)}{\cal Q}_{i+m_1}G_{i+m_1}^{(\ell_2,r_2)-1}.
\]
>From this it follows pairwise compatibility of transformations $s_{m_1}$ and $s_{m_2}$
for any integers $m_1$ and $m_2$. So we can write $s_{m_1}\circ s_{m_2} =
s_{m_2}\circ s_{m_1}$. The inverse maps $s_m^{-1}$ are well-defined by
the formula ${\cal Q}_{i-\ell n+r} = G_{i-\ell n+r}^{(\ell, r)-1}{\cal Q}_iG_{i-\ell n+r}^{(\ell, r)} =
G_i^{(-\ell,-r)}{\cal Q}_iG_i^{(-\ell,-r)-1}$.
$\Box$

Rewrite second equation in (\ref{system1}) as $z^{n-1}H_i\Psi_i(t, z) =
\Psi_{i+n}(t, z) = (\Lambda^n\Psi)_i$.
>From this we derive
\[
z^{k(1-n)}(\Lambda^{kn}\Psi)_i = H_{i+(k-1)n}...H_{i+n}H_i\Psi_i,
\]
\[
z^{k(n-1)}(\Lambda^{-kn}\Psi)_i = H_{i-kn}^{-1}...H_{i-2n}^{-1}H_{i-n}^{-1}\Psi_i.
\]
These relations make one-to-one connection between difference operators
\[
P = \sum_{k\in{\bf Z}}z^{k(1-n)}p_k(t)\Lambda^{kn}\in{\cal D}^{(n,0)}
\]
and sequences
of $\partial^{(n)}$-pseudo-differential operators $\{{\cal P}_i,\; i\in{\bf Z}\}$
mapping the upper triangular part of given matrix (including
main diagonal) into the differential parts of ${\cal P}_i$'s
and the lower triangular part of the matrix to the purely
pseudo-differential parts. More exactly, we have
$(P\Psi)_i = {\cal P}_i\Psi_i$, $(P_{-}\Psi)_i = ({\cal P}_i)_{-}\Psi_i$ and
$(P_{+}\Psi)_i = ({\cal P}_i)_{+}\Psi_i$,
where
\[
{\cal P}_i =
\sum_{k > 0}p_{-k}(i, t)H_{i-kn}^{-1}...H_{i-2n}^{-1}H_{i-n}^{-1} +
\sum_{k\geq 0}p_{k}(i, t)H_{i+(k-1)n}...H_{i+n}H_{i}
\]
\[
= ({\cal P}_i)_{-} +
({\cal P}_i)_{+}.
\]
In what follows, we denote $\sigma : P\in{\cal D}^{(n,0)} \rightarrow
\{{\cal P}_i,\; i\in{\bf Z}\}$. It is easy to check that
\begin{equation}
\sigma : z^{\ell(1-n)}\Lambda^{\ell n - r}Q^{r}
\leftrightarrow \{G_i^{(\ell,r)},\:
i\in{\bf Z}\}.
\label{put}
\end{equation}

{\bf Proposition 5.}
{\it Equations $z^{p(n-1)}\partial_p^{(n)}\Psi = Q_{+}^{pn}\Psi,\;p=2, 3,...$
are equivalent to $\partial_p^{(n)}\psi_i = ({\cal Q}_i^p)_{+}\psi_i,\;p=2, 3,...$.
}

{\bf Proof.} Setting $r=\ell n$ in (\ref{put}) gives
\[
\sigma : z^{p(1-n)}Q^{pn} \leftrightarrow \{{\cal Q}_i^p, i\in{\bf Z}\}.
\]
Taking into account this, we have
\[
z^i\partial_p^{(n)}\psi_i = \partial_p^{(n)}\Psi_i = z^{p(1-n)}(Q_{+}^{pn}\Psi)_i =
({\cal Q}_i^p)_{+}\Psi_i =
z^i({\cal Q}_i^p)_{+}\psi_i.
\]
The latter proves proposition. $\Box$

We learn from this proposition that $n$th discrete KP in fact is equivalent
to bi-infinite sequence of differential KP hierarchies whose evolution equations can
be written as Sato -- Wilson equations
\begin{equation}
\partial_p^{(n)}\hat{w}_i =  ({\cal Q}_i^p)_{+}\hat{w}_i - \hat{w}_i\partial^{(n)p}
\label{SW1}
\end{equation}
where $\hat{w}_i$'s are connected by relations (\ref{system2}) or equivalently
as Lax equations
\begin{equation}
\partial_p^{(n)}{\cal Q}_i = [({\cal Q}_i^p)_{+}, {\cal Q}_i]
\label{LAX2}
\end{equation}
where ${\cal Q}_i$'s are connected by the gauge transformations (\ref{similarity1}).

Let us establish equations managing $G_i^{(r)}$-evolutions with respect to KP
flows. Differentiating l.h.s. and r.h.s. of (\ref{system2}), by virtue
(\ref{SW1}), formally leads to evolution equations
\begin{equation}
\partial_p^{(n)}G_{i}^{(r)} = ({\cal Q}_{i+n-r}^p)_{+}G_{i}^{(r)} - G_{i}^{(r)}({\cal Q}_i^p)_{+}.
\label{LAX3}
\end{equation}
Notice that in the case $r=n$, the latter becomes Lax equations (\ref{LAX2}).
Using standard arguments, one can show that equations (\ref{LAX3}) are properly defined
individually. Indeed, taking into account (\ref{similarity1}), one can
write
$
{\cal Q}_{i+n-r}^p = G_i^{(r)}{\cal Q}_i^pG_i^{(r)-1}
$
or
$
{\cal Q}_{i+n-r}^pG_i^{(r)} = G_i^{(r)}{\cal Q}_i^p
$
for any $p\in{\bf N}$. It follows from this that
\[
({\cal Q}_{i+n-r}^p)_{+}G_i^{(r)} - G_i^{(r)}({\cal Q}_i^p)_{+} =
G_i^{(r)}({\cal Q}_i^p)_{-} - ({\cal Q}_{i+n-r}^p)_{-}G_i^{(r)}.
\]
Thus r.h.s. of (\ref{LAX3}) as well as l.h.s. is a $\Psi$DO
of zero order. Moreover, in the case $r=0$, i.e. when $G_i^{(0)} =
H_i$, r.h.s. of (\ref{LAX3}) is zeroth order differential operator or simply
function. It is easy now to establish $G_i^{(\ell,r)}$-evolutions with
respect to KP flows. This states following proposition.

{\bf Proposition 6.} {\it By virtue of (\ref{recurrence}) and (\ref{LAX3}),
we have
\begin{equation}
\partial_p^{(n)}G_{i}^{(\ell,r)} = ({\cal Q}_{i+\ell n-r}^p)_{+}G_{i}^{(\ell,r)} -
G_i^{(\ell,r)}({\cal Q}_i^p)_{+}.
\label{LAX4}
\end{equation}
}

{\bf Proof.} In the case $\ell=0$, we obtain
\[
\partial_p^{(n)} G_i^{(0,r)}  = \partial_p^{(n)}(G_{i-n}^{(r)}H_{i-n}^{-1}) =
\{({\cal Q}_{i-r}^p)_{+}G_{i-n}^{(r)} - G_{i-n}^{(r)}({\cal Q}_{i-n}^p)_{+}\}H_{i-n}^{-1}
\]
\[
- G_{i-n}^{(r)}H_{i-n}^{-1}\{({\cal Q}_i^p)_{+}H_{i-n} -
H_{i-n}({\cal Q}_{i-n}^p)_{+}\}H_{i-n}^{-1}
\]
\[
=({\cal Q}_{i-r}^p)_{+}G_{i}^{(0,r)} - G_{i}^{(0,r)}({\cal Q}_i^p)_{+}.
\]
Since $G_i^{(1,r)} = G_i^{(r)}$, equation (\ref{LAX4}) in the case $\ell=1$
immediately follows from (\ref{LAX3}). The proof of (\ref{LAX4}) proceeds
by induction. Assume
that (\ref{LAX4}) is valid for some $\ell$, then
\[
\partial_p^{(n)} G_i^{(\ell+1,r)} = \partial_p^{(n)}(G_{i+n}^{(\ell,r)}H_i) =
\{({\cal Q}_{i+(\ell+1)n-r}^p)_{+}G_{i+n}^{(\ell,r)} -
G_{i+n}^{(\ell,r)}({\cal Q}_{i+n}^p)_{+}\}H_i
\]
\[
+ G_{i+n}^{(\ell,r)}\{({\cal Q}_{i+n}^p)_{+}H_i - H_i({\cal Q}_i^p)_{+}\}
\]
\[
= ({\cal Q}_{i+(\ell+1)n-r}^p)_{+}G_i^{(\ell+1,r)} - G_i^{(\ell+1,r)}({\cal Q}_i^p)_{+}.
\]
This proves (\ref{LAX4}) for positive integers $\ell$. By similar
arguments equation (\ref{LAX4}) is showed for negative $\ell$.
$\Box$

{\bf Proposition 7.} {\it Equations (\ref{LAX4}) are pairwise compatible.}

{\bf Proof.}
One must to show that permutation relation (\ref{permutation})
is invariant with respect to KP flows. With the identities
$\ell_1+\ell_2 = \ell_3 + \ell_4$ and $r_1+r_2 = r_3 + r_4$, we have
\[
\partial_p^{(n)} (G_{i+\ell_2n-r_2}^{(\ell_1, r_1)}G_i^{(\ell_2, r_2)})
\]
\[
= \{({\cal Q}_{i+(\ell_1+\ell_2)n-r_1-r_2}^p)_{+}G_{i+\ell_2n-r_2}^{(\ell_1, r_1)} -
G_{i+\ell_2n-r_2}^{(\ell_1, r_1)}({\cal Q}_{i+\ell_2n-r_2}^p)_{+}\}G_i^{(\ell_2, r_2)}
\]
\[
+ G_{i+\ell_2n-r_2}^{(\ell_1, r_1)}\{({\cal Q}_{i+\ell_2n-r_2}^p)_{+}G_i^{(\ell_2, r_2)} -
G_i^{(\ell_2, r_2)}({\cal Q}_i^p)_{+}\}
\]
\[
= ({\cal Q}_{i+(\ell_1+\ell_2)n-r_1-r_2}^p)_{+}G_{i+\ell_2n-r_2}^{(\ell_1, r_1)}G_i^{(\ell_2, r_2)} -
G_{i+\ell_2n-r_2}^{(\ell_1, r_1)}G_i^{(\ell_2, r_2)}({\cal Q}_i^p)_{+}
\]
\[
= ({\cal Q}_{i+(\ell_3+\ell_4)n-r_3-r_4}^p)_{+}G_{i+\ell_3n-r_3}^{(\ell_4, r_4)}G_i^{(\ell_3, r_3)} -
G_{i+\ell_3n-r_3}^{(\ell_4, r_4)}G_i^{(\ell_3, r_3)}({\cal Q}_i^p)_{+}
\]
\[
= \{({\cal Q}_{i+(\ell_3+\ell_4)n-r_3-r_4}^p)_{+}G_{i+\ell_3n-r_3}^{(\ell_4, r_4)} -
G_{i+\ell_3n-r_3}^{(\ell_4, r_4)}({\cal Q}_{i+\ell_3n-r_3}^p)_{+}\}G_i^{(\ell_3, r_3)}
\]
\[
+ G_{i+\ell_3n-r_3}^{(\ell_4, r_4)}\{({\cal Q}_{i+\ell_3n-r_3}^p)_{+}G_i^{(\ell_3, r_3)} -
G_i^{(\ell_3, r_3)}({\cal Q}_i^p)_{+}\} =
\partial_p^{(n)} (G_{i+\ell_3n-r_3}^{(\ell_4, r_4)}G_i^{(\ell_3, r_3)}).
\]
Therefore we proved that equations (\ref{LAX4}) are pairwise consistent.
$\Box$

The fact that $\psi_i(t, z) = \hat{w}_i e^{\xi(t^{(n)}, z)}$ are KP wave
eigenfunctions force they to be expressible via $\tau$-functions
\[
\psi_i(t, z) =
\frac{\tau_i^{(n)}(t^{(1)},..., t^{(n)} - [z^{-1}],...)}
{\tau_i^{(n)}(t^{(1)},..., t^{(n)},...)}
e^{\xi(t^{(n)}, z)}
\]
where $[z^{-1}]\equiv (1/z, 1/(2z^2),...)$.
Define $\Phi_i^{(n)} = \Phi_i^{(n)}(t)$ via $H_i\Phi_i^{(n)} = 0$ or equivalently
through following relation:
\[
\partial^{(n)} \Phi_i^{(n)} = \Phi_i^{(n)}\sum_{s=1}^na_0^{(n)}(i+s-1).
\]
Taking into consideration the second equation in (\ref{LAX3}),
we get
\[
\partial_p^{(n)}(H_i\Phi_i^{(n)}) = ({\cal Q}_{i+n}^p)_{+}H_i\Phi_i^{(n)} -
H_i({\cal Q}_{i}^p)_{+}\Phi_i^{(n)} + H_i\partial_p^{(n)}\Phi_i^{(n)} = 0.
\]
>From this we derive $\partial_p^{(n)}\Phi_i^{(n)} =
({\cal Q}_{i}^p)_{+}\Phi_i^{(n)} +
\alpha_i\Phi_i^{(n)}$ where $\alpha_i$'s are some constants.
Commutativity condition $\partial_p^{(n)}\partial_q^{(n)}\Phi_i^{(n)} =
\partial_q^{(n)}\partial_p^{(n)}\Phi_i^{(n)}$ leads to evolution equations
for KP eigenfunctions $\partial_p^{(n)}\Phi_i^{(n)} =
({\cal Q}_{i}^p)_{+}\Phi_i^{(n)}$,
i.e. $\alpha_i = 0$. Thus the relations
${\cal Q}_{i+n} = H_i{\cal Q}_iH_i^{-1}$ defines DB
transformations with eigenfunctions
$\Phi_i^{(n)} = \tau_{i+n}^{(n)}/\tau_i^{(n)}$ \cite{aratyn}.
It should perhaps to recall that an eigenfunction of
Lax operator ${\cal Q}$ contains information about DB transformation
$\tau \rightarrow \overline{\tau} = \Phi\tau$ while the identity
\[
\{\tau(t - [z^{-1}]), \overline{\tau}(t)\} +
z(\tau(t - [z^{-1}])\overline{\tau}(t) - \overline{\tau}(t - [z^{-1}])
\tau(t)) = 0
\]
with $\{f, g\}\equiv \partial f\cdot g - \partial g\cdot f$ holds.

{\bf Proposition 8.} {\it We have
\begin{equation}
a_{\ell}^{(n, r)}(i) = \frac{p_{\ell+1}(\tilde{D}_{t^{(n)}})\tau_{i-\ell n+r}^{(n)}\circ\tau_{i}^{(n)}}
{\tau_{i-\ell n+r}^{(n)}\tau_{i}^{(n)}}.
\label{coeff}
\end{equation}
}

{\bf Proof.} To show (\ref{coeff}), one need in the well known identity\cite{dickey2}
\begin{equation}
{\rm res}_z[(Pe^{xz})\cdot(Qe^{-xz})] = {\rm res}_{\partial}PQ^{*}
\label{identity1}
\end{equation}
where $P = \sum_{k\in{\bf Z}}p_k(x)\partial^k$ and
$Q = \sum_{k\in{\bf Z}}q_k(x)\partial^k$ are two arbitrary $\Psi$DO's and
$Q^{*}$ is the formal adjoint to $Q$.

To use further the identity (\ref{identity1}) we set
$P = G_i^{(\ell,r)}\hat{w}_i$ and $Q = \hat{w}_i^{* -1}$.
Taking into account (\ref{res}), (\ref{identity1}) and applying
Proposition 2, we get
\[
a_{\ell}^{(n,r)}(i+\ell n-r) =
{\rm res}_z[(G_i^{(\ell,r)}\hat{w}_ie^{x^{(n)}z})(\hat{w}_i^{*-1}e^{-x^{(n)}z})]
\]
\[
= {\rm res}_z[(G_i^{(\ell,r)}\hat{w}_ie^{\xi(t^{(n)},z)})
(\hat{w}_i^{*-1}e^{-\xi(t^{(n)},z)})]
\]
\[
= {\rm res}_z[(G_i^{(\ell,r)}\psi_i(t, z))
\psi_i^{*}(t, z)]
= {\rm res}_z[z^{\ell}\psi_{i+\ell n-r}(t, z)\psi_i^{*}(t, z)]
\]
\[
= {\rm res}_z\left[z^{\ell}
\frac{
\tau_{i+\ell n-r}^{(n)}(t^{(1)},..., t^{(n)} - [z^{-1}],...)
\tau_i^{(n)}(t^{(1)},..., t^{(n)} + [z^{-1}],...)
}
{
\tau_{i+\ell n-r}^{(n)}(t^{(1)},..., t^{(n)},...)
\tau_i^{(n)}(t^{(1)},..., t^{(n)},...)
}
\right]
\]
\[
=
\left.
\frac{1}{(\ell+1)!}\left(\frac{d}{du}\right)^{\ell+1}\left[
\frac{
\tau_{i+\ell n-r}^{(n)}(t^{(1)},..., t^{(n)} - [u],...)
\tau_i^{(n)}(t^{(1)},..., t^{(n)} + [u],...)
}
{
\tau_{i+\ell n-r}^{(n)}(t^{(1)},..., t^{(n)},...)
\tau_i^{(n)}(t^{(1)},..., t^{(n)},...)
}
\right]\right|_{u=0}
\]
Now using technical identity (\ref{identity}) we obtain
\[
a_{\ell}^{(n, r)}(i+\ell n-r) =
\frac{p_{\ell+1}(\tilde{D}_{t^{(n)}})\tau_i^{(n)}\circ\tau_{i+\ell n-r}^{(n)}}
{\tau_i^{(n)}\tau_{i+\ell n-r}^{(n)}}.
\]
Shifting $i\rightarrow i-\ell n+r$ in the latter we arrive at (\ref{coeff}). $\Box$

{\bf Remark 6.} By (\ref{coeff}), we can express $Q^r$ in terms of
$\tau$-functions as
\begin{equation}
Q^r = \sum_{\ell=0}^{\infty}
{\rm diag}\left(
\frac{p_{\ell}(\tilde{D}_{t^{(n)}})\tau_{i-(\ell-1)n+r}^{(n)}\circ\tau_i^{(n)}}
{\tau_{i-(\ell-1)n+r}^{(n)}\tau_i^{(n)}}
\right)_{i\in{\bf Z}}z^{\ell(n-1)}\Lambda^{r-\ell n}.
\label{Q^r}
\end{equation}
In the case of ordinary discrete KP hierarchy ($n=1$), (\ref{Q^r}) coincides
with the formula (0.13) of the paper \cite{adler1}.

Since $a_{\ell}^{(n,0)}(i) = 0$, then as consequence of the above proposition
we deduce following bilinear equations:
\begin{equation}
p_{\ell+1}(\tilde{D}_{t^{(n)}})\tau_{i-\ell n}^{(n)}\circ\tau_i^{(n)} = 0,\;\;
\ell = 0, 1,...
\label{bilinear}
\end{equation}
With the well known bilinear identity for KP
wave eigenfunction (see, for example \cite{dickey})
\[
{\rm res}_z[(\partial_1^{k_1}...\partial_m^{k_m}\psi(t, z))\cdot
\psi^{*}(t, z)] = 0
\]
and the fact that $G_i^{(\ell, \ell n)}, \ell = 0, 1,...$ are
purely differential operators, one can deduce the following.

{\bf Propositon 9.} {\it $\tau$-functions of $n$th discrete
KP hierarchy satisfy
\[
{\rm res}_z[z^{\ell}\tau_{i+\ell n}^{(n)}(t^{(1)},..., t^{(n)}-[z^{-1}],...)
\]
\[
\times
\tau_i^{(n)}(t^{(1)},..., t^{(n)'}+[z^{-1}],...)
\exp\xi(t^{(n)} - t^{(n)'}, z)] = 0,\;\;
\forall t^{(n)}, t^{(n)'}, \;\;
\]
\[
\ell = 0, 1, 2...
\]
}

\section*{Acknowledgments}

This work supported in part by INTAS grant 2000-15.


\section*{References}

\begin{thebibliography}{10}

\bibitem{ueno}
Ueno K and Takasaki K 1984
Toda lattice hierarchy
{\it Adv. Studies in Pure Math.}
{\bf 4}
1-95

\bibitem{date}
Date E, Kashiwara M, Jimbo M and Miwa T 1983
Transformation groups for soliton equations,
{\it Proc. RIMS Simp. Nonlinear Integrable Systems -- Classical
Theory and Quantum Theory (Kyoto)} ed M Jimbo and T Miwa
(Singapore: World Scientific)
pp 39-119

\bibitem{jimbo}
Jimbo M and Miwa T 1983
Solitons and infinite dimensional Lie algebras
{\it Publ. RIMS Kyoto Univ.}
{\bf 19}
943-1001

\bibitem{ohta}
Ohta Y, Satsuma J, Takahashi D and Tokihiro T 1988
An elementary introduction to Sato theory
{\it Prog. Theor. Phys. Suppl.}
{\bf 94}
210-241

\bibitem{dickey2}
Dickey L A 1991
Soliton equations and Hamiltonian systems
{\it Advanced Series in Mathematical Physics}
{\bf 12}
(Singapore: World Scientific)
pp 1-310

\bibitem{adler1}
Adler M and van Moerbeke P 1999
Vertex operator solutions to the discrete KP-hierarchy
{\it Comm. Math. Phys.}
{\bf 203}
185-210

\bibitem{bonora}
Bonora L A and Xiong C S 1992
An alternative approach to KP hierarchy in matrix models
{\it Phys. Lett. B}
{\bf 285}
191-?

\bibitem{aratyn}
Aratyn H, Nissimov E and Pacheva S 1997
Constrained KP hierarchies: additional symmetries,
Darboux-B\"acklund solutions and relations to multi-matrix
models
{\it Int. J. Mod. Phys.}
{\bf A12}
1265-1340

\bibitem{dickey1}
Dickey L A 2001
Chains of KP, semi-infinite 1-Toda lattice hierarchy
and Kontsevich integral, {\it Preprint} nlin.SI/0010054

\bibitem{dickey}
Dickey L A 1999
Modified KP and discrete KP
{\it Lett. Math. Phys.}
{\bf 48}
277-289

\bibitem{svinin}
Svinin A K 2001
$n$th discrete KP hierarchy
{\it Proc. Fourth Conf. Symmetry in Nonlinear Mathematical Physics}
ed A M Nikitin (Kiev: Institute of Mathematics) pp ?-?;
{\it Preprint} nlin.SI/0107049

\bibitem{kuper}
Kupershmidt B A 1985
Discrete Lax equations and differential-difference calculus
(Ast\'erisque)

\bibitem{adler}
Adler V E, Shabat A B and Yamilov R I 2000
Symmetry approach to integrability problem
{\it Teor. Mat. Fiz.}
{\bf 125}
355-424 (in Russian)

\bibitem{bogoyavlenskii}
Bogoyavlenskii O I 1991
Algebraic constructions of integrable dynamical systems ---
extensions of Volterra system
{\it Russian Math. Surveys}
{\bf 46}
1-46

\bibitem{blaszak}
Blaszak M and Marciniak K 1994
$r$-matrix approach to lattice integrable systems
{\it J. Math. Phys.}
{\bf 35}
4661-4682

\bibitem{belov}
Belov A A and Chaltikian K D 1993
Analogues of $W$-algebras and classical integrable equations
{\it Phys. Lett. B}
{\bf 309}
268-274

\bibitem{svinin1}
Svinin A K 2001
A class of integrable lattices and KP hierarchy
{\it J. Phys. A: Math. Gen.}
{\bf 34}
?-?;
{\it Preprint} nlin.SI/0107054

\bibitem{wu}
Wu Y-T and Hu X-B 1999
A new integrable differential-difference system and its
explicit solutions
{\it J. Phys. A: Math. Gen.}
{\bf 32}
1515-1521

\bibitem{wu}
Hu X-B and Tam H-W 2001
New integrable differential-difference systems:
Lax pairs, bilinear forms and soliton solutions
{\it Inverse Problems}
{\bf 17}
319-327

\end{thebibliography}
\end{document}